\documentclass[preprint,notoc]{JHEP} 
\usepackage{epsfig}

 \def\eqn{\begin{equation}}
  \def\eeqn{\end{equation}} \def\arr{\begin{array}} \def\earr{\end{array}}
\def\eqna{\begin{eqnarray}} \def\eeqna{\end{eqnarray}} \def\a{\alpha}
\def\b{\beta} \def\D{\Delta} \def\s{\sigma} \def\d{\delta} \def\w{\wedge}
\def\o{\omega}  \def\O{\Omega}  
\def\th{\theta} \def\m{\mu} \def\n{\nu}  \def\La{\Lambda}
   \def\g{\gamma}
\font\mybb=msbm10 at 12pt \def\bb#1{\hbox{\mybb#1}} 
\def\bR {\bb{R}}

\newcommand\fverb{\setbox\pippobox=\hbox\bgroup\verb}
\newcommand\fverbdo{\egroup\medskip\noindent%
  \fbox{\unhbox\pippobox}\ }
\newcommand\fverbit{\egroup\item[\fbox{\unhbox\pippobox}]} \newbox\pippobox

\title{The Ka\l u\.{z}a-Klein Melvin Solution in M-theory}

\author{Miguel S. Costa\\
  Laboratoire de Physique Th\'eorique de l'\'Ecole Normale Sup\'erieure\\ 
  24 rue Lhomond, F-75231 Paris Cedex 05, France\\
  E-mail: \email{miguel@lpt.ens.fr}}

\author{Michael Gutperle\\
  Jefferson Laboratory of Physics\\
  Harvard University, Cambridge, MA 02138, USA\\
  E-mail: \email{gutperle@riemann.harvard.edu}}

\preprint{\hepth{0012072}\\LPTENS-00-44\\HUTP-00-A048}

\abstract{We study some aspects of the Ka\l u\.{z}a-Klein Melvin
solution in M-theory. The associated magnetic field has a maximal
critical value $B=\pm 1/R$ where $R$ is the radius of the
compactification circle. It is argued that the Melvin background of
type IIA with magnetic field $B$ and of type 0A with magnetic field
$B'=B-1/R$ are equivalent. Evidence for this conjecture is provided
using a further circle compactification and a `9-11' flip. We show that 
partition functions of nine-dimensional type IIA strings and of a  
$(-1)^F\sigma_{1/2}$ type IIA orbifold both with NS-NS Melvin
fluxtubes are related by such shift of the magnetic field. 
Then the instabilities of both IIA and 0A Melvin
solutions are analyzed. For each theory there is an instanton
associated to the decay of spacetime.  In the IIA case the decay mode
is associated to the nucleation of $D6/D\bar{6}$-brane pairs, while in
the 0A case spacetime decays through Witten's bubble production.}

\begin{document} 

\section{Introduction}

The study of string theories in strong background fields has received 
considerable attention over the past years. For example, in the
context of the AdS/CFT duality, the near horizon limit of D3-branes gives IIB
string theory in an $AdS_5\times S_5$ background with self-dual
R-R 5-form flux \cite{maldacena}. More recently, open strings and membranes 
in near critical electric fields were seen to produce new interesting
dynamical regimes in string theory and M-theory 
\cite{seiberg,gopakumara,gopakumarb,bergshoeff}.    

In this paper some aspects of another example of a string background
with nontrivial flux will be discussed. The Melvin fluxtube universe
\cite{melvin} describes an axisymmetric spacetime with a
non-vanishing magnetic flux along the axis. Interestingly such a
spacetime can be realized as the Ka\l u\.{z}a-Klein reduction from a
flat higher dimensional spacetime with nontrivial identifications. In
a nice series of papers \cite{dowkera,dowkerb,dowkerc,dowkerd}, Dowker
et al. analyzed the Ka\l u\.{z}a-Klein Melvin solution. In particular,
they discussed the nonperturbative instability and pair production of
magnetically charged black holes. In this work we apply some of their
results in the M-theory context.

The plan of this paper is as follows. In section two we describe the
Ka\l u\.{z}a-Klein  Melvin solution of IIA strings coming from dimensional
reduction of M-theory.  In particular, we discuss the appearance of a
maximal magnetic field and the issue of spin structures. The nonzero
magnetic flux in the Ka\l u\.{z}a-Klein Melvin solution breaks all the
supersymmetries. We shall argue that this Melvin solution is equivalent
to 0A string theory \cite{dixon,seibergb} in a (different) magnetic
field. A `9-11' flip relates these two theories to the nine-dimensional
type II strings and to the $S_1/(-1)^F\s_{1/2}$ type IIA orbifold both
with a NS-NS Melvin fluxtube. In this case the
equivalence of both theories can be analyzed explicitly.

In section three the instabilities of the Melvin background are
discussed. These are
associated to instantons describing decay modes of the
spacetime. There are two instantons, namely the shifted and unshifted
instantons, which differ by their spin structures and give the decay
modes for the Melvin IIA and 0A string backgrounds, respectively. In
the former case the instanton is associated to the nucleation of
$D6/D\bar{6}$-brane pairs while in the latter to a deformed version of
Witten's expanding bubble \cite{wittena}.

We give our conclusions in section four.

\section{Melvin Solution from M-theory}\label{sectwo}

An interesting  solution of Einstein-Maxwell theory is the  Melvin solution
\cite{melvin} which represents a magnetic fluxtube universe.
One of the most surprising features of the Ka\l u\.{z}a-Klein Melvin solution
\cite{Gibbonsa} is that the
corresponding higher dimensional spacetime is flat but has
nontrivial identifications \cite{dowkerb,dowkerc}. Consider
the eleven-dimensional flat metric in M-theory written in cylindrical
coordinates 
\eqn
ds^2=-dt^2+dy_mdy^m+dz^2+d\rho^2+\rho^2d\varphi^2+dx_{11}^{\ \ 2}\ ,
\label{flatmetric}
\eeqn where $m=1,\cdots,6$. For simplicity  the six-dimensional
manifold spanned by $y^m$ is taken to be  a six-torus.
Next make the following identification: 
\eqn
(t,y_m,z,\rho,\varphi,x_{11}) \equiv (t,y_m,z,\rho,\varphi+2\pi n_1RB+2\pi
n_2,x_{11}+2\pi n_1R)\ ,
\label{identifications}
\eeqn 
for every integer $n_1$ and $n_2$. The identifications under the
shifts $2\pi n_1R$ for $x_{11}$ and $2\pi n_2$ for $\varphi$ are
standard. However, the shift  $x_{11}\rightarrow x_{11}+2\pi n_1R$ is
accompanied by a rotation $\varphi\rightarrow \varphi+2\pi n_1RB$. 
Upon dimensional reduction along the Killing vector 
\eqn
l=\partial_{11}+B\,\partial_{\varphi}
\label{Killvector}
\eeqn 
the above identifications give rise to the dilatonic Melvin background
\cite{Gibbonsa}.  It is convenient to
introduce the coordinate $\tilde{\varphi}=\varphi-B\,x_{11}$, which is
constant along orbits of $l$  and has standard periodicity. Using the
relation between the M-theory metric and the string frame metric,
dilaton field and R-R 1-form potential\footnote{Notice for
  future reference that with the factor of $2$ multiplying the R-R 1-form
  potential, the $D6$-brane tension $T_6$ and charge density $\rho_6$ are
  related by $\rho_6=2T_6$.}  
\eqn ds_{11}^{\ \ 2}=e^{-2\phi/3}ds_{10}^{\ \ 2}+ 
e^{4\phi/3}\left(dx_{11}+2{\cal A}_{\m}dx^{\m}\right)^2\ ,
\label{reduction}
\eeqn 
the ten-dimensional IIA background is described by
\eqn \arr{c} \displaystyle{ ds_{10}^{\ \ 
    2}=\La^{1/2}\left(-dt^2+dy_mdy^m+dz^2+d\rho^2\right)
  +\La^{-1/2}\rho^2d\tilde{\varphi}^2\ ,}
\\
\displaystyle{ e^{4\phi/3}=\La \equiv 1+\rho^2B^2\ ,\ \ \ \ {\cal
    A}_{\tilde{\varphi}}=\frac{B\rho^2}{2\La}\ .}  \earr
\label{Melvin}
\eeqn 
The parameter $B$ is the magnetic field along the $z$-axis defined by
$B^2=\frac{1}{2}F_{\m\n}F^{\m\n}|_{\rho=0}$.  Since the eleven-dimensional
spacetime is flat this metric is expected to be an exact solution of the
M-theory including higher derivative terms.\footnote{The Green-Schwarz
  String propagating in the IIA Melvin background was discussed in \cite{tseytlinh}} However, from the IIA
perspective, the Melvin background is always a non-perturbative vacuum
because far away from the axis, i.e. when $\rho$
becomes large, the string coupling becomes large and  the appropriate
description is eleven-dimensional.

For $\rho\ll 1/|B|$ spacetime is approximately flat and the
compactification radius is approximately 
constant. In this region the string coupling is $g=R/\sqrt{\a'}$ and 
perturbative string theory can be trusted provided $R$ is small. 
In addition, for the Ka\l u\.{z}a-Klein ansatz to make sense, 
the length scale $\rho$ must be much larger than the compactification
scale, i.e. $\rho\gg R$. Hence, the condition $|B|\ll 1/R$ is necessary for
this background to be a good  approximation for IIA theory in a
constant R-R magnetic field near $\rho=0$. 

Near the `critical' magnetic field   $|B|\sim 1/R$ the 
region where space is flat and the theory
weakly coupled is much smaller than the compactification radius.
This implies  that 
the ten-dimensional interpretation of the solution is not applicable
anywhere.

\subsection{Spin Structures}\label{secspst}

Since the angle $\varphi$ has period $2\pi$ the identifications
(\ref{identifications}) are not altered by the replacement  $B\rightarrow
B+n/R$. Hence the magnetic field $B$ describes inequivalent spacetimes
within the range \cite{dowkerc}
\eqn 
-\frac{1}{2R} < B \le \frac{1}{2R}\ .
\label{naiverange}
\eeqn 
However, the above argument is not quite correct in a theory which
includes fermions, like M-theory or superstring theory.  In
fact we shall see that physics changes drastically under the shift
$B\rightarrow B+n/R$ for $n$ odd. A fermion acquires a phase equal to
$-1$ under a rotation $\D\varphi=2\pi$ with the identity corresponding 
to a rotation by $4\pi$. Hence, physics is unaltered only when
$\varphi$ is shifted by $4\pi$. From (\ref{identifications}) it
follows that the spacetime identifications and the boundary conditions
on the fermions do not change under the shift $B\rightarrow
B+2n/R$. We conclude that inequivalent spacetimes lie within the range 
\eqn 
-\frac{1}{R} < B \le \frac{1}{R}\ .
\label{range}
\eeqn

More formally the Melvin solution has topology 
${\cal M}^6\times\bR^4\times S^1$, where ${\cal M}^6$ is the topology
of the compact six-dimensional space which  we shall neglect. In what
follows we shall denote the Cartesian coordinates in the $\rho,\phi$
plane of (\ref{flatmetric}) by $x_1,x_2$. The manifold 
$\bR^4\times S^1$ admits two different spin structures. 
A vector parallelly  
transported around the $S^1$ will be rotated by an angle
$\D\varphi=2\pi RB$, while a fermion parallelly transported around the $S^1$
will return with a phase 
\eqn 
e^{\pi RB \gamma_{12}} \ \ \ {\rm or}\ \ \ 
-e^{\pi RB \gamma_{12}}\ ,
\label{phases}
\eeqn 
according to the spin structure.  Here $\g_{12}$ is a generator of
the $Spin(3,1)$ Lie algebra which generates a rotation in the
$x_{1,2}$-plane. We neglect the action of the spin group
$G$ due to parallel transport along the compact six-dimensional
manifold. Since $(\gamma_{12})^2=-1$ it is easy to see that for
$B\rightarrow B+1/R$ the spin structures are interchanged while for
$B\rightarrow B+2/R$ we obtain the same physics.

Although the eleven-dimensional spacetime is flat, the nontrivial
identification implies that for $B\neq 0$ no Killing spinors exist, i.e.
\ from the ten-dimensional point of view supersymmetry will be broken by the
presence of the magnetic field. Related  approaches to supersymmetry
breaking include the Scherk-Schwarz mechanism \cite{Scherk,Kounnasa} 
and magnetized tori \cite{bachas}. 
Remarkably, from the type IIA string point of view the argument
given above predicts the existence of a maximum magnetic field
associated with the R-R 1-form potential  
\eqn
|B_{max}|=\frac{1}{R}=\frac{1}{g\sqrt{\a'}}\ .
\label{maxB}
\eeqn 
This is a non-perturbative prediction and it is a geometrical
consequence of the Ka\l u\.{z}a-Klein reduction of M-theory to the IIA
strings.  It is tempting to conjecture that this is the maximum magnetic
field associated with a R-R 1-form potential that can be confined to a
region of spacetime in string theory. Also, we expect that under
T-duality a similar result holds for the other R-R $p$-form fields. We
think that the existence of a maximum electric/magnetic field ought to
have a deep explanation. In fact, for a NS-NS magnetic field on a
3-sphere the same phenomenon occurs \cite{Kounnasb}. In this
case the physical interpretation of the maximum magnetic field is
clear, at the critical value the states that couple to the magnetic
field become infinitely massive and decouple. It would be interesting
to understand better this issue. 

\subsection{Type 0A Strings}

The Ka\l u\.{z}a-Klein Melvin background admits two different
spin structures as described in (\ref{phases}). For $B=0$, the
first case in (\ref{phases}) corresponds to the supersymmetric IIA
compactification of M-theory (fermions obey periodic boundary
conditions on $S^1$), while the second case corresponds to the
non-supersymmetric 0A compactification of M-theory (fermions
obey anti-periodic boundary conditions on $S^1$) \cite{berga}. As
explained in section \ref{secspst},  under the shift $B\rightarrow
B+1/R$ the spin structures are inverted, therefore it is natural to
suspect that the IIA and 0A theories are related by such shift of the
magnetic field. For example, the IIA theory with critical field B=1/R
corresponds to the flat 11D space with the spin structure inverted.
Reducing along the Killing vector
$l=\partial_{11}+\frac{1}{R}\,\partial_{\varphi}$ gives the IIA
theory in a critical magnetic field, which is strongly coupled
everywhere. Equivalently, reducing along the Killing vector
$l'=\partial_{11}$ and adopting the second spin structure in
(\ref{phases}) gives the 0A theory with vanishing magnetic field.

To be more precise consider the IIA theory with maximum magnetic
$B=1/R$. The radius $R$ is related to the string coupling $g$ by
$R=g\sqrt{\a'}$. It is important to realize that this is the coupling
constant at $\rho=0$. As explained before, the condition for the
coupling to remain small $\rho\ll 1/B=R$ implies that
the theory is in the eleven-dimensional
regime (although we are keeping $g$ small). From the 0A point of view
the situation is rather different: the magnetic field is zero and the
string coupling $g'$ satisfies $g'=R/(2\sqrt{\a'})=g/2$ which is
small. Hence, the appropriate description of the IIA theory at critical
magnetic field, which is strongly coupled, is given by the weakly
coupled 0A theory.

More generally, consider the first spin structure in (\ref{phases})
and reduce the spacetime described by (\ref{flatmetric}) along 
\eqn
l=\partial_{11}+B\,\partial_{\varphi}
\eeqn
to obtain the IIA theory in a magnetic field $B$. This is equivalent
to adopt the second choice in (\ref{phases}) and reduce along
\eqn
l'=\partial_{11}+\left(B\mp\frac{1}{R}\right)\partial_{\varphi}
\eeqn 
to obtain the 0A theory in a magnetic field $B'=B\mp\frac{1}{R}$,
where the $\mp$ choice correspond to $B>0$ and $B<0$, respectively.
Hence, we conclude that the IIA theory in a magnetic field $B$ is dual
to the 0A theory in a magnetic field $B'=B\mp\frac{1}{R}$, with the
couplings at the $z$-axis related by $R=g\sqrt{\a'}=2g'\sqrt{\a'}$. 
Notice that the Melvin solution for the 0A theory is similar to the
IIA solution (\ref{Melvin}) with the R-R 1-form potential in the
untwisted sector. The tachyon and twisted R-R fields are set to zero.

\subsection{NS-NS Melvin Fluxtube}

Because quantization of string theories in R-R backgrounds is not well
understood it is difficult to directly prove our conjecture. One can however
relate the IIA Melvin background to a background of perturbative string
theory  by further compactifying the IIA theory with a R-R fluxtube
on a circle and by using a `9/11 flip'. The resulting theory will be the
nine-dimensional type II theory with a NS-NS fluxtube. 

Applying the construction of section \ref{sectwo} to IIA string theory
in ten dimensions relates the flat ten-dimensional spacetime with metric
\eqn
ds^2=-dt^2+dy_mdy^m+dz^2+d\rho^2+\rho^2d\varphi^2+dx_{9}^{\ \ 2}\ ,
\label{flatmetricten}
\eeqn
to a nine-dimensional magnetic fluxtube. Here $m=1,\cdots,5$ and
$x_{11}$ is replaced by $x_9$ in the nontrivial identification
(\ref{identifications}). The Ka\l u\.{z}a-Klein reduction  
\eqn 
ds_{10}^{\ \ 2}=ds_9^{\ 2}+
e^{2\sigma}\left(dx_{9}+2{\cal A}_{\m}dx^{\m}\right)^2\ ,
\label{tendkk}
\eeqn
gives the NS-NS Melvin solution 
\eqn 
\arr{c} 
\displaystyle{ds_9^{\ 2}=-dt^2+dy_mdy^m+dz^2 +d\rho^2 
+\La^{-1}\rho^2 d\varphi^2\ ,}
\\
\displaystyle{A_\varphi = \frac{B \rho^2}{2\La}\ ,\quad
\quad e^{2\sigma}\equiv\La= 1+B^2\rho^2\ .} 
\earr 
\label{nsnsmel}
\eeqn 
The gauge field $A_\varphi$ is a NS-NS field, coming from the metric
element $g_{\mu9}$ under Ka\l u\.{z}a-Klein reduction. Note that 
the nine-dimensional dilaton field $\phi_9=\phi-\sigma/2$ is not
constant. Since the ten-dimensional dilaton field
is trivial and no R-R fields are turned on, it is
straightforward to calculate the string spectrum and the partition
function of this background. Although the ten-dimensional background
is flat the identifications imply: Firstly  the string fields in the
$x_1,x_2$-directions are twisted due to the nontrivial rotation in the
$x_{1,2}$-plane; Secondly the momentum and winding modes in the
$x_9$-direction have an additional contribution depending on the
angular momentum in the $x_{1,2}$-plane.

It can be shown that the quantization of bosonic strings
\cite{tseytlinc} and superstrings  \cite{tseytlina,tseytlinb}  moving
in the  background (\ref{nsnsmel}) can be
reduced to free fields. For the superstring 
the mass spectrum is given by 
\eqn
\arr{rcl}
mass^2&=&\displaystyle{{2\over \alpha^\prime}(N_L+N_R)
+{m^2\over R^2}+ w^2{R^2\over {\alpha^\prime}^2}-2B{m\over R}(J_L+J_R)}\\
&&\displaystyle{-2B{Rw\over
\alpha^\prime}(J_L-J_R)+ B^2 (J_L+J_R)^2}\ ,
\earr
\label{massf}
\eeqn 
where $m,w$ are the momentum and winding number of the $X_9$ circle
of radius $R$ and $J_L,J_R$ are the left- and right-moving angular
momentum operators in the $x_{1,2}$-plane, respectively. The spectrum
is non-supersymmetric for $B\neq 2n/R$ and it contains tachyons coming
\ from the winding sector $w\neq 0$ for $R<\sqrt{2\alpha^\prime}$ and 
$B\le B_{crit}=R/(2\a')$. The one loop partition function was also
derived in \cite{tseytlina,tseytlinb} and is of the form 
\eqn 
\arr{rcl}
Z(B,R)&=&\displaystyle{V_7 R 
\int {d^2\tau\over \tau_2^{\ 2}} {1\over\tau_2^{\ 3}}
\sum_{m,n}\exp\left(-{\pi R^2\over\alpha^\prime\tau_2} 
\left| m+n\tau\right|^2\right)}
\\
&&\displaystyle{\left|\theta
\left[\arr{c} 1+BRn\\ 1+BRm\earr\right](0|\tau)\right|^8\;
{1\over\left|\theta_1^\prime(0|\tau)\right|^6
\;\left|\theta
\left[\arr{c} 1+2BRn\\ 1+2BRm\earr\right]
(0|\tau)\right|^2}\ ,} 
\earr
\label{partftw}
\eeqn 
where 
$\theta\left[\arr{c} 1\\ 1\earr\right](v|\tau)=\theta_1(v|\tau)$. 
In the Green-Schwarz formalism the structure of the partition function
becomes very transparent since  the twisting of the eight
Green-Schwarz fermions and the two bosons in the $x_1,x_2$-directions
are responsible for the twisting of the theta functions in the
numerator and denominator of (\ref{partftw}), respectively.

The properties of theta functions ensure that the partition function
$Z(B,R)$ is invariant under $B\to B+2/R$ and hence the spectrum
(\ref{massf}) is invariant under this shift. The partition function
vanishes when $B=2n/R$ with $n\in Z$, indicating that the spectrum is
the one of the ten-dimensional IIA string compactified on a circle.

For the critical value of the magnetic field $B=(2n+1)/R$ the spectrum is
equivalent to the superstring with anti-periodic boundary conditions for
the spacetime fermions \cite{rohma,attikwitten,greenf}. Note that for these values
translation invariance is restored in the $x_{1,2}$-plane which
introduces a (divergent) volume factor, coming from the twisted
bosons. Then the partition function reads 
\eqn 
\arr{rcl} 
Z_{crit}=Z(B,R)\mid_{B={1\over R}}&=&
\displaystyle{V_9 R\int{d^2\tau\over \tau_2^{\ 2}}{1\over \tau_2^{\ 3}}
\left\{Z_{0,0}
{\left|\theta_3^{\ 4}(0|\tau)-\theta_4^{\ 4}(0|\tau)
-\theta_2^{\ 4}(0|\tau)\right|^2 
\over\left|\eta(\tau)\right|^{24}}\right.}
\\
\\
&&\displaystyle{\left.+Z_{1,0} 
{\left|\theta_2(0|\tau)\right|^8\over\left|\eta(\tau)\right|^{24}}
+Z_{0,1}
{\left|\theta_4(0|\tau)\right|^8\over\left|\eta(\tau)\right|^{24}} 
+Z_{1,1}
{\left|\theta_3(0|\tau)\right|^8\over\left|\eta(\tau)\right|^{24}}
\right\}\ ,} 
\earr
\label{partftwb}
\eeqn 
where $Z_{ab}$ defines the summation over odd and even winding
numbers around the two cycles of the torus 
\eqn 
Z_{ab}(\tau,R)=\sum_{m,n}
\exp\left(-{\pi R^2\over\alpha^\prime\tau_2}
\mid(2m+a)+(2n+b)\tau\mid^2\right)\ .
\label{defzab}
\eeqn
A Poisson resummation on $m$ of the partition function
$Z_{ab}$ defined in (\ref{defzab}) yields
\eqn
Z_{ab}(\tau,R)= {\sqrt{\alpha^\prime \tau_2}\over 2R}\sum_{kl}
(-1)^{ak} \exp\left(-\pi\tau_2\left[{\alpha^\prime\over 4R^2}k^2+
{R^2\over \alpha^\prime}(2l+b)^2\right] +\pi i \tau_1 k(2l+b)\right)\ .
\label{poisonrs}
\eeqn
Defining the standard  $SO(8)$ characters as follows 
\eqn
\chi_o= {1\over \eta(\tau)^{12}}
\left(\theta_3^{\ 4}+\theta_4^{\ 4}\right),\quad
\chi_v= {1\over \eta(\tau)^{12}}
\left(\theta_3^{\ 4}-\theta_4^{\ 4}\right),\quad 
\chi_{s,c}= {1\over \eta(\tau)^{12}}\theta_2^{\ 4}\ ,
\label{chardef}
\eeqn
it is straightforward to show that the partition function
(\ref{partftwb}) can be expressed in the following way 
\eqn 
\arr{c}
\displaystyle{Z_{crit}= V_9 R  \int 
{d^2\tau\over \tau_2^{\ 2}} {1\over \tau_2^{\ 3}}\Big\{
\big(\chi_v\bar\chi_v+\chi_s\bar\chi_s
-\chi_v\bar\chi_s-\chi_s\bar\chi_v\big)Z_{00}}
\\
\displaystyle{-\big(\chi_v\bar\chi_v+\chi_s\bar\chi_s
+\chi_v\bar\chi_s+\chi_s\bar\chi_v\big) Z_{10}+\big
(\chi_o\bar\chi_o+\chi_c\bar\chi_c-\chi_o\bar\chi_c
-\chi_c\bar\chi_o\big)Z_{01}}
\\
\displaystyle{-\big(
\chi_o\bar\chi_o+\chi_c\bar\chi_c+\chi_o\bar\chi_c
+\chi_c\bar\chi_o\big)Z_{11}\Big\}}\ .
\earr
\label{Zorb}
\eeqn
The partition function for the critical Melvin solution (\ref{Zorb}) is
the same as for the type IIA $S_1/(-1)^F\sigma_{1/2}$ 
orbifold with the identification $R_{orb}=2R_{Melvin}$. Here $F$
is the spacetime fermion number and $\sigma_{1/2}$ is a shift by half the
circumference of the circle. The first two terms in (\ref{Zorb}) come
\ from the projection onto invariant states and the last two terms are
coming from the twisted sector states by modular invariance. 

The type IIA $S_1/(-1)^F\sigma_{1/2}$ orbifold interpolates between
type IIA and 0A theories in the limits $R\to \infty$ and $R\to 0$, 
respectively. In \cite{berga} it was argued that exchanging the
orbifold and M-theory circle (`9/11 flip') the ten-dimensional 0A
theory is related to a $S_1/(-1)^F\sigma_{1/2}$ orbifold of M-theory. 
This `9/11 flip' will map the NS-NS fluxtube to a R-R fluxtube,
showing that the ten-dimensional 0A theory is related to the IIA
theory at the critical R-R magnetic field, as conjectured earlier.

The relation of the type II NS-NS Melvin solution with magnetic field
$B$ and Ka\l u\.{z}a-Klein radius $R$, and the type IIA 
$S_1/(-1)^F\sigma_{1/2}$ orbifold with magnetic field $B'=B-1/R$
(for $B>0$) and radius $R'=2R$ can be seen using the Lagrangian
representation of the Melvin solution (\ref{partftw}). Then,
one can split the sum over $m,n$ into four sectors
corresponding to $(m,n)$ being $(e,e),(o,e),(e,o),(o,o)$, these four
sectors correspond to the untwisted sector $(1,1)$ and the three
twisted sectors $(g,1),(1,g),(g,g)$ of the $g= (-1)^F\sigma_{1/2}$
orbifold. 

\section{Instabilities}

Schwinger pair production is mapped under electromagnetic duality to pair
production of magnetically charged particles in a constant magnetic
field. In Yang-Mills-Higgs theories magnetic monopoles can be pair produced
\cite{mantona,mantonb}, similarly, in a gravitational theory
magnetically charged black holes can be pair produced 
\cite{dowkera,dowkerb,dowkerc,dowkerd,gibbons,stromingera,stromingerb,rossa}.
For the IIA R-R-Melvin background we expect the  production of
$D6/D\bar{6}$-brane pairs \cite{dowkerd}. Also, we expect some decay mode for
the 0A theory. Within the supergravity approximation to M-theory
standard semiclassical quantum gravity instanton methods 
\cite{HawGibb} can be applied   
to estimate the rate for the decay process\footnote{The
evaluation of the one loop determinants in quantum
gravity is problematic and will not be discussed here.}.
All we have to do is to find the instanton associated with the process 
and calculate its action. Then the rate for the nucleation process is
estimated by $\Gamma\sim e^{-I}$. The subsequent Lorentzian evolution
is determined by a surface of zero extrinsic curvature on the
Euclidean manifold to which a Lorentzian manifold is glued.

\subsection{Type 0A and IIA Instantons}

The instanton associated with the Ka\l u\.{z}a-Klein
monopole/anti-monopole pair production is the Myers-Perry Kerr
instanton \cite{myers} with a single (complexified) angular momentum
parameter and six extra flat directions \cite{dowkerc}. The metric for this 
background reads
\eqn
\arr{rcl}
ds^2&=&\displaystyle{
dy_mdy^m+dx_{11}^{\ \ 2}+\sin^2{\th}\left(r^2-\a^2\right)d\varphi^2
-\frac{\m}{\rho^2}\left(dx_{11}+\a\sin^2{\th}\,d\varphi\right)^2}
\\
&&\displaystyle{
+\frac{\rho^2}{r^2-\a^2-\m}\,dr^2+\rho^2d\th^2+r^2\cos^2{\th}\,d\psi^2\ ,}
\earr
\label{instanton}
\eeqn
where $\rho^2=r^2-\a^2\cos^2\th$. There is a coordinate singularity
at $r^2=r_H^{\ \,2}\equiv\m+\a^2$, the locus of the black hole horizon in the
Lorentzian metric. The parameters of the background (\ref{instanton})
can be expressed in terms of 
\eqn
B=\frac{\a}{\m}\ ,\ \ \ \ 
\frac{1}{R}=k=\frac{\sqrt{\m+\a^2}}{\m}\ ,
\label{unshifted}
\eeqn
where  $\o\equiv i\O=iB$ and $k$ 
are the Lorentzian angular velocity and 
surface gravity, respectively. 

We shall reduce the above manifold along the orbits  of the vector
$l=\partial_{11}+B\,\partial_{\varphi}$, which
has zero norm at $r=r_H$. As in the case of the 
 Melvin solution this can be done 
by changing to the coordinate 
$\tilde{\varphi}=\varphi-B\,x_{11}$ 
and reducing along $l=\partial_{11}$. 
Then, the conical singularity at $r=r_H$
can be removed provided $x_{11}$ has period $2\pi R$ for fixed 
$\tilde{\varphi}$. In the limit $r\rightarrow \infty$ the instanton
metric becomes flat. Because of the identifications on the angles
$x_{11}$ and $\varphi$ the instanton approaches the Euclidean
Ka\l u\.{z}a-Klein Melvin solution describing a decay mode of this
background.

From the relations (\ref{unshifted}) it follows  that the 
parameter $B$ lies within, $-\frac{1}{R}\le B \le \frac{1}{R}$. 
This means that a 
Kerr instanton with $B>1/(2R)$ ($B<1/(2R)$) asymptotically approaches the 
same magnetic field as the same instanton with parameter $B-1/R$ 
($B+1/R$). Hence we can define the shifted Kerr instanton with 
parameters \cite{dowkerc}
\eqn
B=\frac{\a}{\m}\mp\frac{1}{R}\ ,\ \ \ \ 
\frac{1}{R}=k=\frac{\sqrt{\m+\a^2}}{\m}\ ,
\label{shifted}
\eeqn
where the $\mp$ signs corresponds to $\a>0$ and $\a<0$, respectively.
Both unshifted and shifted instantons (\ref{unshifted}) and 
(\ref{shifted}) approach asymptotically the same magnetic field.
However, the Melvin backgrounds with magnetic field $B$ and 
$B\mp1/R$ have different spin structures. 
This is reassuring because
these two instantons are not identical, allowing the appropriate
identification of an instanton as the decay mode of an associated
vacuum with given spin structure. In other words, since there are
two distinct Kerr instantons approaching asymptotically the same 
magnetic field the appropriate decay mode is fixed 
by the spin structure. In fact, both instantons
are the same Kerr instanton with different values of its parameters
but their ten-dimensional interpretation is quite different.

Since the Kerr instanton has topology $\bR^2\times S^3$, which
is connected, it admits a single spin structure. Spinors
parallelly transported around an integral curve of $l$ pick
a phase $-e^{\pi R\frac{\a}{\m}\g}$ for both unshifted and shifted
instantons \cite{dowkerc}. If we choose the first spin
structure for the Melvin background in (\ref{phases}), corresponding
to the IIA theory, the appropriate Kerr instanton  is the
shifted one where $\frac{\a}{\m}=B\mp\frac{1}{R}$. On the
other hand, the second choice in (\ref{phases}) picks the unshifted 
instanton as the decay mode for the  0A theory.
Remarkably, the spin structures uniquely fix the spacetime decay modes
over the physically inequivalent values of the magnetic field 
for both IIA and 0A theories.

The decay probability rate can be estimated as $e^{-I}$ where
the Euclidean action for the instantons reads
\eqn
I=-\frac{1}{16\pi G_{11}}\int d^{11}x\,\sqrt{g}\,R
-\frac{1}{8\pi G_{11}}\int d^{10}x\,\sqrt{h}\left(K-K_0\right)\ .
\label{action}
\eeqn
$K$ is the trace of the extrinsic curvature of the boundary with
metric $h_{ij}$ and $K_0$ is the same quantity for the Melvin background.
For the unshifted Kerr instanton we have \cite{dowkerc}
\eqn
I=\frac{\pi V_6}{8G_{10}}\frac{R^2}{1-(BR)^2}\ ,
\label{unshiftedaction}
\eeqn
while for the shifted instanton the action is
\eqn
I=\frac{\pi V_6}{8G_{10}}\frac{R^2}{1-\left(1-|B|R\right)^2}\ ,
\label{shiftedaction}
\eeqn
where $V_6$ is the volume of the compact six-dimensional manifold.
The ten-dimensional Newton constant $G_{10}$ satisfies 
$16\pi G_{10}=(2\pi)^7g^2\a'^4$. In figure 1 the action for both
instantons as a function of the magnetic field is ploted. 

\FIGURE{\epsfig{file=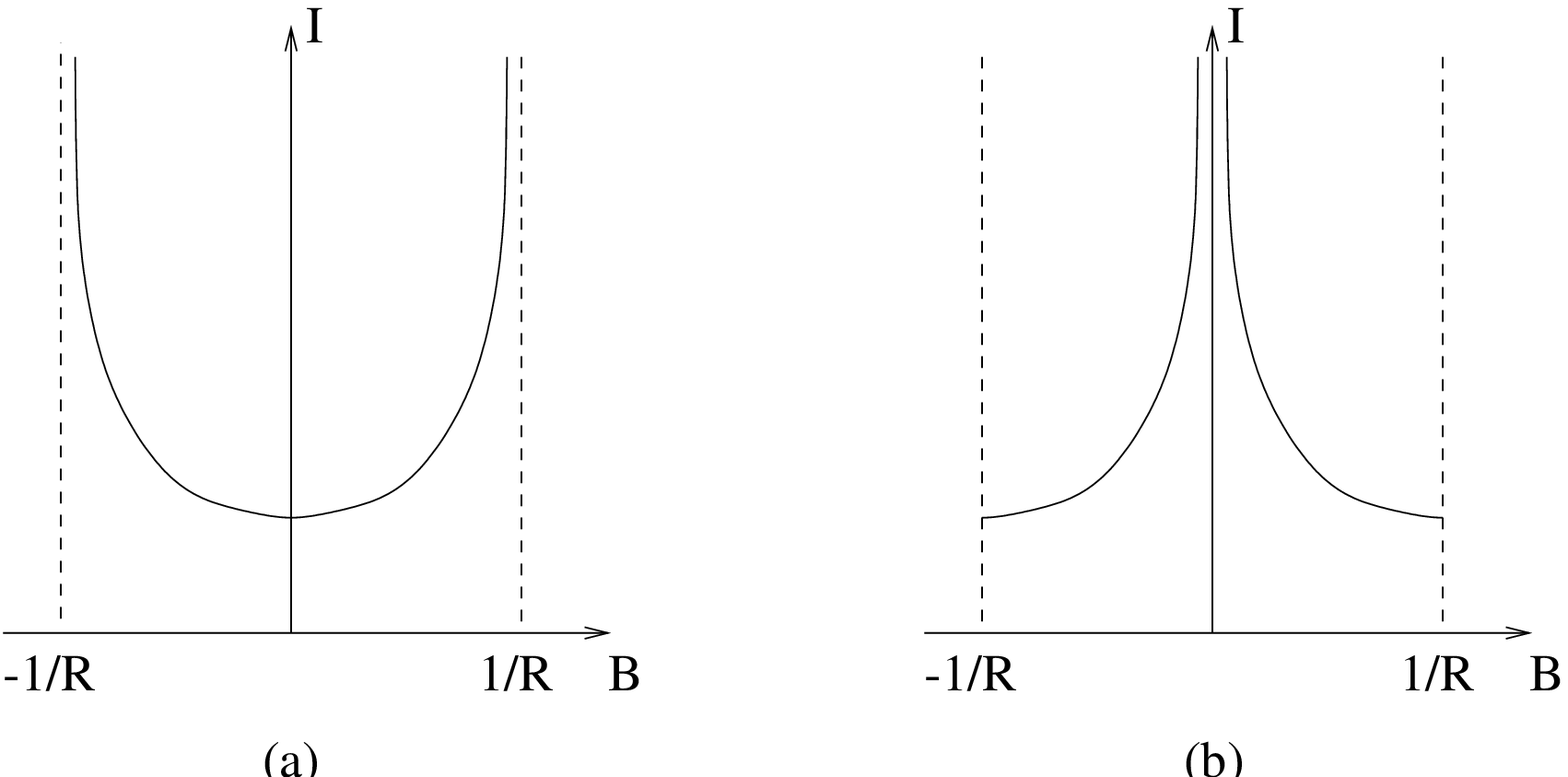,width=5in}
        \caption{\small{The actions for the instantons associated with the
decay of the Melvin background for (a) the 0A theory and (b) the IIA
theory. In the former case the decay is through Witten's bubble
production while in the latter through
$D6/D\bar{6}$-brane pair production. The actions go to
infinity for $B=\pm 1/R$ for the unshifted instanton and for $B=0$ for the
shifted instanton, in agreement with stability of supersymmetric
type IIA strings.}}}

\subsection{Decay Modes}

The Euclidean instanton is associated to a tunneling trajectory
describing the decay of a spacetime. The post decay evolution depends
on a surface of zero extrinsic curvature in the Euclidean
manifold. This surface provides the initial data to obtain a
Lorentzian metric describing the spacetime into which the Melvin
background decays. Such a surface is defined by $\psi=0$ and
$\psi=\pi$ with $0\le\th\le\pi/2$, which is joined at $\th=\pi/2$. The
subsequent Lorentzian evolution is obtained by analytic continuation
$\psi\rightarrow it$. From the eleven-dimensional point of view
the interpretation for both instantons is the same. It describes an
expanding bubble without spherical symmetry \cite{dowkerc}, 
i.e. it is a deformation
of Witten's expanding bubble \cite{wittena} (an expanding cigar). The
expanding two-surface $r=r_{H}$ has initial area $4\pi\m$ and expands
like $\cosh^2{t}$. In the case of the 0A theory without magnetic field
(or the IIA theory with critical field) the decay mode is indeed given
by Witten's expanding bubble corresponding to the Euclidean
Schwarzchild instanton.

\subsubsection{Type 0A}

Next we describe the decay process as seen by a ten-dimensional observer. 
Firstly consider the unshifted instanton associated with the 0A theory.
We reduce to ten-dimensions along the vector
\eqn
l=\partial_{11}+B\,\partial_{\varphi}
=\partial_{11}+\O\,\partial_{\varphi}\ ,
\label{unshreduction}
\eeqn
where $\O=\a/\m$ is the Euclidean angular velocity. Because the reduction 
is co-rotating with the geometry, a single singularity appears at
$r=r_{H}$ where $l^2=0$. Of course this is an artifact of the 
reduction since the geometry is perfectly regular
\ from the eleven-dimensional point of view. The subsequent Lorentzian
evolution is that of a space falling into the singularity \cite{dowkerc}.

It is interesting to consider the instanton action for vanishing
magnetic field. This will be associated to the bubble nucleation
instability of the type 0A theory. The instanton action reads
\eqn
I=\frac{\pi V_6}{8G_{10}}R^2=
4\pi \frac{V_6}{(2\pi)^6\alpha'^3}\ ,
\label{Schw}
\eeqn
which does not depend on the
ten-dimensional string coupling $g'$. In the case that
$V_6\ll\alpha'^3$, i.e. when the theory is effectively
four-dimensional, one might think that the decay of spacetime through
bubble production is dominant. In this case the length scale $R$ is of
order of the Planck length and the semi-classical approximation is no
longer valid. However, we still expect this to indicate an
instability. In fact, we can consider the T-dual theory by dualizing
the 6-torus. Then the false vacuum is long lived and the
semi-classical calculation reliable.

\subsubsection{Type IIA}

Secondly consider the shifted instanton associated with the IIA theory.
Now we reduce along the orbits of the vector
\eqn
l=\partial_{11}+B\,\partial_{\varphi}
=\partial_{11}+\left(\O\mp\frac{1}{R}\right)\partial_{\varphi}\ .
\label{shreduction}
\eeqn
The horizon $r=r_{H}$ becomes a line horizon where the metric
is regular except for the two singularities at the ends ($\th=0,\pi$)
\cite{dowkerc}. This line horizon becomes a cigar from the 11D
perspective. Subsequently the singularities will accelerate apart
\ from each other (corresponding to the ends of an expanding cigar in 11D).

If  $|B|\ll1/R$ the Melvin solution describes the perturbative IIA
theory in a constant magnetic R-R field within a region $\rho\ll R\ll
1/|B|$. Hence it is natural to ask whether the creation of
$D6/D\bar{6}$-brane pairs as described by semiclassical gravity
follows the expected behavior for the Schwinger pair production. 
In the limit  $|B|\ll1/R$ the action (\ref{shiftedaction}) becomes 
\eqn
I=\frac{\pi V_6}{8G_{10}}\frac{R}{2|B|}
=\frac{\pi M_6}{2|B|}\ ,
\label{schwinger}
\eeqn
where $M_6=V_6T_6$ and  $T_6=((2\pi)^6\a'^{7/2}g)^{-1}$ is the 
$D6$-brane tension. This gives exactly the Schwinger nucleation rate 
$\Gamma\sim e^{-\pi\frac{m^2}{qE}}$ for a particle with charge $q=2m$ 
in a constant electric field $E$. In fact, the $D6$-brane action in the 
perturbative IIA theory region is for vanishing worldvolume gauge
field (notice that in our conventions $\rho_6=2T_6$)
\eqn
S=-T_6\int d^7\s~ e^{-\phi}\sqrt{-\det{\hat{G}_{\a\b}}}
-2T_6\int \hat{{\cal A}}_7\ ,
\label{BI}
\eeqn
where $\hat{G}_{\a\b}$ and $\hat{{\cal A}}_7$ are the pull-backs to the
worldvolume of the metric and dual R-R 7-form potential
($\star{\cal F}_2=d{\cal A}_7$), respectively. A short calculation shows
that in the Melvin background
\eqn
d{\cal A}_7=B~dt\w dz\w dy_1\w\cdots\w dy_6\ .
\label{electric}
\eeqn
This indicates that the instanton calculation gives exactly the expected
result for a BPS saturated $D6$-brane subjected to a constant 
electric field of magnitude $E=B$. 

The magnetically charged objects
which will be created are $D6$-$\bar D6$ pairs  wrapped on the $T^6$.
Hence, from the four-dimensional point of view these are D0-branes
which couple electrically to the dual of the magnetic field. Then the
action of the instanton can be easily reproduced in a simple probe
calculation of the D0 brane in the Melvin background.
Assume that the D-particle moves only in the (Euclidean) two plane spanned
by $\tau$ and $z$ and that the only non-vanishing component of the
electromagnetic tensor is $F_{\tau z}=B$. The Born Infeld action for
the Euclidean signature reads
\eqn
S= M_6 \int d\s
\left(\sqrt{{\dot \tau}^2+{\dot z}^2 }-B z\dot\tau\right)\ .
\eeqn
The equations of motion following from this action are
\eqn
{\partial\over \partial \s}
\left( {\dot\tau\over\sqrt{{\dot\tau}^2+{\dot z}^2}}-Bz\right)=0\ ,
\quad\quad  
{\partial\over \partial \s}
\left( {\dot z\over\sqrt{{\dot \tau}^2+{\dot z}^2}}+B\tau\right)=0 \ .
\label{eqofm}
\eeqn
The integration constant can be set to zero by shifting $\tau, z$ and a
solution of these equations is given by 
\eqn
\tau ={1\over B} \sin{\s}\ ,\quad\quad z={1\over B}\cos{\s}\ .
\label{actionb}
\eeqn
This solution parameterizes a circle of radius $1/B$ in the range 
$\s\in [0,2\pi]$. Evaluating the action gives
\eqn
S={M_6\over |B|}\int_{0}^{2\pi}d\s
\left(1-\cos^2{\s}\right)={\pi M_6 \over 2|B|}\ , 
\label{actiond}
\eeqn
which agrees exactly with (\ref{schwinger}).

\subsection{Non-compact Spacetime}

So far our discussion considered a ten-dimensional spacetime where six
dimensions live on a $T^6$, i.e. we focused on the
compactification to four dimensions. However, the
results above reported can be generalized to theories with a
non-compact ten-dimensional spacetime. We shall describe briefly such
generalization and refer the reader to \cite{dowkerd} for the
details. 

In the case of a non-compact type IIA Melvin universe described by 
(\ref{Melvin}) the instanton action (\ref{shiftedaction}) for the 
$D6/D\bar{6}$ pair production is infinite because the volume $V_6$ 
is infinite. However, the Melvin vacuum can decay via the nucleation of
a single spherical $D6$-brane. The appropriate instanton is
the Myers-Perry eleven-dimensional shifted instanton with a single 
(complexified) angular momentum. For small magnetic field 
$|B|\ll 1/R$, the action agrees with that of a spherical probe with 
action given by (\ref{BI}) calculated in the perturbative IIA region.
At critical magnetic field $|B|=1/R$, the ten-dimensional type IIA 
description breaks down and the appropriate description is given by
 weakly coupled type 0A theory. In this case we consider the
eleven-dimensional unshifted Schwarzschild instanton that describes the
bubble (a 8-sphere) nucleation in the type 0A vacuum. The action is
\eqn
I=\frac{2^{10}\O_7}{G_{10}}R^8=\frac{2^{15}\O_7}{\pi^6}g'^6\ ,
\eeqn
where $\O_7$ is the unit 7-sphere volume. As for the compact spacetime
there are two important regimes. When the compactification radius is
large compared to the Planck length the false vacuum is long lived and
spacetime will eventually decay through bubble production. This
happens at the strong coupling regime where the 0A tachyon is expected
to become massive \cite{berga}. When the compactification radius is of
the order of the Planck length the semi-classical approximation breaks
down. At this point one expects the perturbative 0A tachyon to appear
in the spectrum signaling the instability of the theory. 
The relation of nonperturbative instabilities and the appearance of
tachyons in the perturbative spectrum has been discussed in several
other contexts \cite{Callan:1998kz,Fabinger:2000jd,Horowitz:2000gn}.  

\section{Conclusion}

For the type IIA theory, the typical distance $\d$ between the
nucleated brane and anti-brane
can be estimated by associating the area $A=4\pi\m$ of the cigar to
the initial mass of a membrane connecting the monopole/anti-monopole
pair. The membrane mass is given by $M=A\,T_m=\d\,T_s$, where $T_m$ is
the membrane tension related to the string tension $T_s$ by 
$T_s=2\pi R\,T_m$. For $|B|\ll1/R$ we have $A=\pi R/|B|$ and the $D6/D\bar{6}$
pair is created at a typical distance $\d\sim 1/(2|B|)\gg R$. 
On the other hand, in the limit of critical magnetic field $|B|=1/R$
the decay mode for the IIA theory is given by the shifted
Schwarzschild instanton. In this case $A=4\pi R^2$ and the mass for
strings ending on the $D6/D\bar{6}$-branes is
$M=2g\sqrt{\a'}\,T_s$. From the perturbative strings point of view
this classical mass vanishes and would give the usual string tachyon
after quantization. However, for larger values of $g$ this tachyon
acquires a positive classical mass which is associated with a minimal
distance between the $D6/D\bar{6}$ pair given by 
$\d\sim 2g\sqrt{\a'}$. As explained before, for the IIA theory at
critical magnetic field the appropriate weakly coupled description is
given by the type 0A theory with vanishing magnetic field and string
coupling $g'=g/2$. In this case spacetime decays through bubble
nucleation and the string tachyon becomes massive at strong coupling.

Finally we comment on the analysis by Sen \cite{sena} of the
four-dimensional Kerr instanton with one flat time direction and six
flat spatial directions. After reduction this solution is interpreted
as a $D6/D\bar{6}$-brane pair kept in precarious equilibrium in a R-R
magnetic field. Sen interpreted the case of a critical magnetic field
$|B|=1/R$ as a coincident brane/anti-brane pair because reducing along
a different direction the critical field can be brought to a vanishing
magnetic field. In the light of the discussion above, the appropriate
weakly coupled description is given by the type 0A theory. The mass of
the string tachyon ending on the $D6/D\bar{6}$ pair is now interpreted
as the mass of the 0A tachyon, in the same spirit of the previous
paragraph.

In conclusion, we have uncovered a relation between the Melvin
background and type IIA and 0A strings. This connection arose as a
consequence of the M-theory Ka\l u\.{z}a-Klein reduction to IIA and 0A
strings. An immediate consequence of the dimensional reduction is the
existence of a maximum magnetic field. Then a careful analysis of the
spacetime spin structures led us to conjecture that under a shift of
the Melvin magnetic field both IIA and 0A theories are
equivalent. Evidence for this relation was given by using perturbative
string theory and the `9/11' flip symmetry of M-theory. We proceeded
by analyzing the instabilities of both IIA and 0A
Melvin solutions. In each case and for physically inequivalent
magnetic fields there is an instanton associated to the decay of
spacetime. In the case of the IIA theory this instanton is associated
to the nucleation of $D6/D\bar{6}$-brane pairs, while in the 0A case
to Witten's expanding bubble. 

\acknowledgments
We wish to thank C. Bachas, L. Cornalba and C. Kounnas for helpful comments.
The work of M.G. was  supported in part by the David and Lucile Packard
Foundation. The work of M.S.C was supported by a Marie Curie
Fellowship under the European Commission's Improving Human Potential
programme.

\end{document}